\documentstyle[psfig]{mn}

\def\etal{{\it et al.\ }}
\def\eg{{\it e.g.\ }}

\def\spose#1{\hbox to 0pt{#1\hss}}
\def\approxlt{\mathrel{\spose{\lower 3pt\hbox{$\sim$}}
	\raise 2.0pt\hbox{$<$}}}
\def\approxgt{\mathrel{\spose{\lower 3pt\hbox{$\sim$}}
	\raise 2.0pt\hbox{$>$}}}
\def\approxpropto{\mathrel{\spose{\lower 3pt\hbox{$\sim$}}
	\raise 2.0pt\hbox{$\propto$}}}
\mathchardef\twiddle="2218

\def\multleft#1{\hbox to size{\vbox {\halign {\lft{##}\cr #1}}\hfill}\par}
\def\multright#1{\hbox to size{\vbox {\halign {\rt{##}\cr #1}}\hfill}\par}

\def\today{\ifcase\month\or January\or February\or March\or April\or May\or
      June\or July\or August\or September\or October\or November\or December\fi
      \space\number\day, \number\year}
\def\<{\thinspace}

\def\km{{\rm\thinspace km}}

\def\Mpc{{\rm\thinspace Mpc}}

\def\s{{\rm\thinspace s}}
\def\yr{{\rm\thinspace yr}}

\def\kmps{\hbox{$\km\s^{-1}\,$}}

\def\kmpspMpc{\hbox{$\kmps\Mpc^{-1}$}}

\title[cooling flows and metallicity measurements]
{The relationship between cooling flows and metallicity measurements for
X-ray luminous clusters.}
\author[S.W. Allen and A.C Fabian]
{\parbox[]{6.in} {S.W. Allen and A.C. Fabian  \\
\footnotesize
Institute of Astronomy, Madingley Road, Cambridge CB3 OHA\\
}}
\begin{document}
\maketitle

\begin{abstract}
We explore the relationship between the metallicity of the intracluster
gas in clusters of galaxies, determined by X-ray spectroscopy, and the
presence of cooling flows. Using ASCA spectra and 
ROSAT images, we demonstrate a clear segregation between the 
metallicities of clusters with and without
cooling flows. On average, cooling-flow clusters have an emission-weighted
metallicity a factor $\sim 1.8$ times higher than that of non-cooling flow 
systems. We suggest this to be due to the presence of metallicity 
gradients in the cooling flow clusters, coupled with the sharply
peaked X-ray surface brightness profiles of these systems. 
Non-cooling flow clusters have much flatter X-ray surface 
brightness distributions and are thought to have undergone recent 
merger events which may have mixed the central high-metallicity gas with the
surrounding less metal-rich material. We find no evidence for evolution
in the emission-weighted metallicities of clusters
within $z \sim 0.3$.

\end{abstract}

\begin{keywords}
galaxies: clusters: general -- cooling flows -- intergalactic medium -- 
X-rays: galaxies
\end{keywords}
\section{Introduction}

X-ray spectroscopy provides an accurate measure of the metallicity of the
hot intracluster medium (ICM) in clusters of galaxies. The strengths of the
various emission lines relative to the continuum reveal the 
abundances of the emitting elements relative to hydrogen. Such measurements
are particularly clear for the case of iron in rich clusters where the
K-shell lines are typically well-defined in the X-ray spectra. The
mass of the ICM dominates over the visible mass in stars in a cluster by
a factor of 2--5 (Arnaud \etal 1992; David, Jones \& Forman 1995; White
\& Fabian 1995). Metallicity measurements from X-ray observations
thus provide firm constraints on the history of metal production within the cluster
potential wells.

For low-redshift clusters, the observed abundance of iron in the ICM 
is approximately 1/3 solar (\eg Edge \& Stewart 1991). 
The strong correlation between the mass of iron and the total 
optical light from elliptical and lenticular galaxies within a cluster suggests that
early-type galaxies are responsible for the bulk of the enrichment (Arnaud
1992). Mushotzky \etal (1996) showed that the 
relative abundances of individual elements such as Si, S and Fe determined from 
X-ray spectra suggest that most of the metals originate from type II supernovae. 
Mushotzky and Lowenstein (1997) further demonstrated that the iron abundance in rich
clusters shows little evolution between $z \sim 0.3$ and now, suggesting
that most of the enrichment of the ICM occurred at high redshifts ($z > 0.3$). 
This is consistent with current semi-analytic models
of galaxy formation (e.g. Kauffmann \& Charlot 1997) which find that more
than 80 per cent of the metal enrichment occurs at $z > 1$.

X-ray observations of clusters of galaxies show that in the central regions
of most ($70-90$ per cent) clusters the cooling time of the ICM is
significantly less than the Hubble time (Edge \etal 1992; 
Peres \etal 1997).  The observed cooling leads to a slow net
inflow of material towards the cluster centre; a process known as a cooling
flow (Fabian 1994). The X-ray imaging data show that gas typically `cools
out' throughout the central few tens to hundreds of kpc in the clusters.
Recent spatially resolved X-ray
spectroscopy has confirmed the presence of distributed cool (and rapidly
cooling) gas in cooling flows, with a spatial distribution and luminosity in
excellent agreement with the predictions from the imaging data (Allen \&
Fabian 1997). The cooling flow in an X-ray luminous cluster can 
account for up to $\sim 70$ per cent of the total bolometric luminosity
of the system (about half of this luminosity being due to material
cooling out of the flow, and the rest due to the gravitational work done on 
the gas as it flows inwards; Allen \etal 1998). 
 
Where abundance measurements have been compiled for large samples of 
clusters, a significant dispersion in the metallicity (which is primarily
determined from the iron abundance) for clusters of a fixed X-ray luminosity and/or temperature 
has been revealed. In particular, the metallicity appears to depend on whether or not a 
cluster has a cooling flow (Yamashita 1992; Fabian \etal 1994a) in the
sense that cooling-flow clusters tend to have higher metallicities. 
Fabian \etal (1994a) speculated that this may be due to differences in 
the origins of the clusters, to the intracluster gas in the cooling-flow clusters 
being inhomogeneous with the cooler gas being more metal-rich, or due to
the presence of metallicity gradients in these systems. Those clusters 
without cooling flows are assumed to have experienced a major merger event which has 
mixed the gas, leading to a lower {\it emission-weighted} metallicity.

In this paper we present metallicity measurements for a sample of 30 X-ray 
luminous clusters observed with ASCA and ROSAT (see Allen \etal 1998 for details). 
We identify the cooling flows in the sample and
explicitly account for the effects of the cooling gas on the X-ray
spectra. We argue that an abundance gradient in the cooling flow 
clusters underlies the variation seen in the emission-weighted
metallicities. Throughout this paper, we assume 
$H_0$=50 \kmpspMpc, $\Omega = 1$ and $\Lambda = 0$.

\section{Observations and data analysis}

\begin{table*}
\vskip 0.2truein
\begin{center}
\caption{Summary of the cooling time, metallicity and baryon fraction measurements} 
\vskip 0.2truein
\begin{tabular}{ c c c c c c c c c c }
\hline
\multicolumn{1}{c}{} &
\multicolumn{1}{c}{} &
\multicolumn{1}{c}{$z$} &
\multicolumn{2}{c}{$t_{\rm cool}$} &
\multicolumn{1}{c}{} &
\multicolumn{2}{c}{Metallicity} &
\multicolumn{1}{c}{} &
\multicolumn{1}{c}{$M_{\rm gas}/M_{\rm total}$ } \\
\multicolumn{1}{c}{} &
\multicolumn{1}{c}{} &
\multicolumn{1}{c}{} &
\multicolumn{2}{c}{($10^9$ yr)} &
\multicolumn{1}{c}{} &
\multicolumn{2}{c}{(solar)} &
\multicolumn{1}{c}{} &
\multicolumn{1}{c}{(at 500 kpc)} \\
\hline                                                               
&&&&&&&& \\
COOLING FLOWS   & ~ &       &   Central         & 100 kpc   & ~ &        MODEL A         &  MODEL C               & ~ &     MODEL C    \\
&&&&&&&& \\						          					                
Abell 478       & ~ & 0.088 & $1.1^{+0.1}_{-0.1}$   & 3.11  & ~ & $0.32^{+0.02}_{-0.03}$ & $0.35^{+0.03}_{-0.04}$ & ~ & $0.22^{+0.03}_{-0.03}$ \\ 
Abell 586       & ~ & 0.171 & $5.5^{+0.7}_{-0.7}$   & 7.94  & ~ & $0.31^{+0.13}_{-0.13}$ & $0.36^{+0.16}_{-0.15}$ & ~ & $0.10^{+0.06}_{-0.05}$ \\ 
PKS0745-191     & ~ & 0.103 & $1.1^{+0.1}_{-0.1}$   & 2.65  & ~ & $0.31^{+0.03}_{-0.03}$ & $0.35^{+0.04}_{-0.03}$ & ~ & $0.20^{+0.03}_{-0.03}$ \\ 
IRAS 09104+4109 & ~ & 0.442 & $2.0^{+0.1}_{-0.1}$   & 2.32  & ~ & $0.43^{+0.13}_{-0.12}$ & $0.51^{+0.17}_{-0.15}$ & ~ & $0.12^{+0.03}_{-0.05}$ \\ 
Abell 963       & ~ & 0.206 & $4.1^{+1.3}_{-0.5}$   & 5.60  & ~ & $0.30^{+0.08}_{-0.07}$ & $0.31^{+0.07}_{-0.13}$ & ~ & $0.21^{+0.03}_{-0.02}$ \\ 
Zwicky 3146     & ~ & 0.291 & $1.7^{+0.1}_{-0.1}$   & 2.53  & ~ & $0.27^{+0.06}_{-0.06}$ & $0.33^{+0.07}_{-0.06}$ & ~ & $0.16^{+0.05}_{-0.05}$ \\ 
Abell 1068      & ~ & 0.139 & $1.2^{+0.1}_{-0.1}$   & 2.75  & ~ & $0.43^{+0.08}_{-0.08}$ & $0.42^{+0.10}_{-0.08}$ & ~ & $0.17^{+0.04}_{-0.03}$ \\ 
Abell 1413      & ~ & 0.143 & $8.6^{+0.6}_{-0.7}$   & 8.63  & ~ & $0.28^{+0.05}_{-0.05}$ & $0.30^{+0.06}_{-0.05}$ & ~ & $0.10^{+0.01}_{-0.01}$ \\ 
Abell 1689      & ~ & 0.184 & $2.9^{+0.5}_{-0.2}$   & 4.34  & ~ & $0.29^{+0.05}_{-0.05}$ & $0.30^{+0.06}_{-0.05}$ & ~ & $0.18^{+0.01}_{-0.02}$ \\ 
Abell 1704      & ~ & 0.216 & $2.1^{+0.1}_{-0.2}$   & 3.59  & ~ & $0.34^{+0.15}_{-0.13}$ & $0.38^{+0.17}_{-0.16}$ & ~ & $0.21^{+0.06}_{-0.08}$ \\ 
RXJ1347.5-1145  & ~ & 0.451 & $2.6^{+0.2}_{-0.1}$   & 3.00  & ~ & $0.38^{+0.11}_{-0.10}$ & $0.43^{+0.11}_{-0.11}$ & ~ & $0.09^{+0.06}_{-0.02}$ \\ 
Abell 1795$^*$  & ~ & 0.063 & $1.4^{+0.1}_{-0.1}$   & 3.42  & ~ & $0.36^{+0.03}_{-0.02}$ & $0.36^{+0.03}_{-0.02}$ & ~ & $0.23^{+0.01}_{-0.01}$ \\ 
MS1358.4+6245   & ~ & 0.327 & $2.8^{+1.2}_{-0.6}$   & 4.76  & ~ & $0.32^{+0.15}_{-0.15}$ & $0.38^{+0.16}_{-0.15}$ & ~ & $0.17^{+0.06}_{-0.08}$ \\ 
Abell 1835      & ~ & 0.252 & $1.5^{+0.5}_{-0.3}$   & 2.44  & ~ & $0.35^{+0.06}_{-0.05}$ & $0.40^{+0.06}_{-0.06}$ & ~ & $0.24^{+0.03}_{-0.04}$ \\ 
MS1455.0+2232   & ~ & 0.258 & $1.1^{+0.2}_{-0.1}$   & 2.15  & ~ & $0.32^{+0.09}_{-0.07}$ & $0.37^{+0.09}_{-0.08}$ & ~ & $0.26^{+0.05}_{-0.09}$ \\ 
Abell 2029      & ~ & 0.077 & $1.5^{+0.1}_{-0.1}$   & 3.54  & ~ & $0.43^{+0.03}_{-0.03}$ & $0.46^{+0.03}_{-0.03}$ & ~ & $0.18^{+0.01}_{-0.02}$ \\ 
Abell 2142      & ~ & 0.089 & $4.3^{+0.8}_{-0.7}$   & 5.38  & ~ & $0.25^{+0.05}_{-0.05}$ & $0.27^{+0.05}_{-0.05}$ & ~ & $0.18^{+0.01}_{-0.03}$ \\ 
Abell 2204      & ~ & 0.152 & $0.94^{+0.04}_{-0.04}$& 2.01  & ~ & $0.41^{+0.06}_{-0.06}$ & $0.46^{+0.07}_{-0.07}$ & ~ & $0.20^{+0.02}_{-0.04}$ \\ 
Abell 2261      & ~ & 0.224 & $3.0^{+1.4}_{-0.6}$   & 5.63  & ~ & $0.32^{+0.10}_{-0.09}$ & $0.37^{+0.09}_{-0.09}$ & ~ & $0.13^{+0.04}_{-0.04}$ \\ 
MS2137.3-2353   & ~ & 0.313 & $1.2^{+0.1}_{-0.1}$   & 1.65  & ~ & $0.44^{+0.14}_{-0.13}$ & $0.50^{+0.15}_{-0.14}$ & ~ & $0.21^{+0.03}_{-0.05}$ \\ 
Abell 2390      & ~ & 0.233 & $4.2^{+0.3}_{-0.2}$   & 5.45  & ~ & $0.36^{+0.15}_{-0.16}$ & $0.40^{+0.18}_{-0.16}$ & ~ & $0.15^{+0.07}_{-0.08}$ \\ 
&&&&&&&& \\								     					                           
\hline                                                               						    
MEAN            & ~ & $0.21\pm0.11$ & $2.61\pm1.89$   & $3.95\pm1.91$  & ~ & $0.344\pm0.057$     &  $0.381\pm0.065$                 & ~ & $0.177\pm0.048$               \\
\hline                                                               
&&&&&&&& \\								     
&&&&&&&& \\								     
NON-COOLING FLOWS  & ~ &       &   Central         & 100 kpc   & ~ &  MODEL A            & & ~ &     MODEL A        \\
&&&&&&&& \\						        			           
Abell 2744      & ~ & 0.308 & $18.8^{+42.0}_{-6.4}$ & 19.2  & ~ & $0.17^{+0.09}_{-0.09}$ & --- & ~ & $0.15^{+0.02}_{-0.01}$ \\ 
Abell 520       & ~ & 0.203 & $16.7^{+24.0}_{-6.2}$ & 16.3  & ~ & $0.14^{+0.10}_{-0.10}$ & --- & ~ & $0.15^{+0.03}_{-0.02}$ \\ 
Abell 665       & ~ & 0.182 & $12.3^{+0.9}_{-0.9}$  & 13.0  & ~ & $0.22^{+0.07}_{-0.08}$ & --- & ~ & $0.12^{+0.01}_{-0.01}$ \\ 
Abell 773       & ~ & 0.217 & $9.8^{+13.7}_{-6.6}$  & 11.2  & ~ & $0.21^{+0.09}_{-0.09}$ & --- & ~ & $0.11^{+0.02}_{-0.02}$ \\ 
Abell 2163      & ~ & 0.208 & $12.7^{+12.3}_{-4.6}$ & 12.1  & ~ & $0.23^{+0.07}_{-0.08}$ & --- & ~ & $0.13^{+0.01}_{-0.01}$ \\ 
Abell 2218      & ~ & 0.175 & $10.4^{+4.1}_{-2.0}$  & 12.8  & ~ & $0.18^{+0.06}_{-0.06}$ & --- & ~ & $0.14^{+0.02}_{-0.02}$ \\ 
Abell 2219      & ~ & 0.228 & $9.0^{+18.8}_{-3.8}$  & 10.7  & ~ & $0.18^{+0.08}_{-0.08}$ & --- & ~ & $0.15^{+0.01}_{-0.02}$ \\ 
Abell 2319      & ~ & 0.056 & $9.5^{+10.4}_{-3.2}$  & 9.81  & ~ & $0.33^{+0.06}_{-0.06}$ & --- & ~ & $0.12^{+0.01}_{-0.01}$ \\ 
AC114           & ~ & 0.312 & $16.9^{+32.7}_{-6.2}$ & 16.6  & ~ & $0.20^{+0.12}_{-0.13}$ & --- & ~ & $0.10^{+0.01}_{-0.02}$ \\ 
&&&&&&&& \\						         			                               						       
\hline                                                               			   
MEAN            & ~ & $0.21\pm0.08$ &  $12.90\pm3.68$    & $13.52\pm3.15$ & ~ & $0.207\pm0.054$                  & & ~ & $0.130\pm0.019$                   \\
\hline                                                               
&&&&&&&& \\						       									       
\end{tabular}															      
\end{center}
\parbox {7in}
{NOTES: Column 2 lists the redshifts for the
clusters. Columns 3 and 4 list the central cooling times and the 
mean cooling times within 100 kpc of the cluster centres determined from
the deprojection analysis of the ROSAT HRI images. 
Columns 5 and 6 summarize the metallicity measurements for the clusters 
from the ASCA data using both the isothermal spectral model (A) and, for
the CF clusters, the more sophisticated model incorporating the cooling-flow emission 
component (model C). Column 7 lists the ratios of the X-ray gas mass to the
total mass at a radius of 500 kpc in the clusters. The errors on the central cooling 
times are the 10 and 90 percentile values determined from 100
Monte Carlo simulations in the deprojection analysis. Errors on the 
metallicity results and baryon fractions are 90 per cent ($\Delta \chi^2
=2.71$) confidence limits. Errors on the mean
values are the standard deviations of the distributions. 
} 
\end{table*}

Full details of the reduction and analysis of the total sample of 
30 clusters are given in Allen et al (1998). ROSAT High Resolution Imager
(HRI) images were used to map the X-ray surface brightness profiles of the 
cluster cores 
and determine whether the individual clusters have cooling flows or not. 
The ASCA spectra were then fitted with a series of appropriate spectral models 
(see below) from which the metallicity measurements were made. 

For the purposes of this paper, we have classified the clusters into
subsamples of cooling-flow (CF) and non-cooling flow (NCF) systems. CFs
are those clusters for which the upper (90 per cent confidence) limit to
the central cooling time, as determined from a deprojection analysis of
the ROSAT HRI X-ray images, is less than $10^{10}$ yr. NCFs
are those systems with upper limits to their central cooling times $>
10^{10}$ yr. [The `central' cooling time is the mean cooling time of 
the cluster gas in the innermost bin included in the deprojection 
analysis, which is of variable size (see Allen \etal 1998 for details). 
The use of a fixed physical size of 100 kpc for the central bin 
leads to very similar results (see Table).]  
Using this simple classification we identify 21 CFs and 9
NCFs in our sample. The mean redshift for the subsamples of
both CF and NCF clusters is $\bar{z} = 0.21$. 

The modelling of the ASCA spectra was carried out using the XSPEC spectral
fitting package (version 9.0; Arnaud 1996). The spectra were modelled
using the plasma codes of Kaastra \& Mewe (1993; incorporating the Fe L
calculations by Liedhal in XSPEC version 9.0) 
and the photoelectric absorption models of Balucinska-Church \&
McCammon (1992). The data from all four ASCA detectors were analysed 
simultaneously with the fit parameters linked to take the same
values across the data sets. The exceptions were the emission
measures of the ambient cluster gas in the four detectors which, due to 
the different extraction radii used, were allowed to fit independently.

Three models were fitted to the spectra. Model A, consisted of an 
isothermal plasma in collisional equilibrium, at the optically-determined redshift for the 
cluster, and absorbed by the nominal Galactic column density (Dickey \& Lockman 1990). 
The free parameters in this model were the temperature ($kT$) and
metallicity ($Z$) of the plasma and the emission measures in the four detectors. (The 
metallicities are determined relative to the solar values of Anders \& Grevesse (1989) 
with the different elements assumed to be present in solar ratios.) 
Secondly, model B, which was identical to model A but with the absorbing column
density ($N_{\rm H}$) also included as a free parameter in the fits.
Thirdly, model C, which included an additional component
explicitly accounting for the emission from the cooling flows in the clusters. The
material in the cooling flows is assumed to cool at constant pressure from
the ambient cluster temperature, following the prescription of Johnstone
\etal (1992). The normalization of the cooling-flow component is
parameterized in terms of a mass deposition rate, ${\dot M}$, which was free
parameter in the fits. (The metallicity of the cooling gas was assumed to
be equal to that of the ambient ICM.) The cooling flows were also assumed to be 
absorbed by an intrinsic column density, $\Delta N_{\rm H}$ (Allen \&
Fabian 1997 and references therein), which was a further 
free parameter in the fits. The metallicity of X-ray absorbing material 
was fixed at
the solar value (Anders \& Grevesse 1989).

\section{The Metallicity measurements}

\begin{figure}
\centerline{\hspace{3cm}\psfig{figure=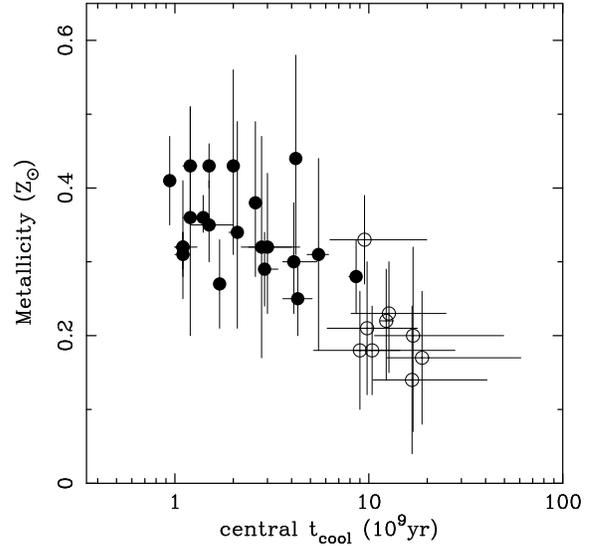,width=0.65\textwidth
,angle=270}}
\caption{The emission-weighted 
metallicity of the cluster gas (determined with spectral model A) 
versus the central cooling time (from the deprojection analysis of the
ROSAT HRI images). CFs are plotted as filled circles and non-cooling flows
as open symbols. The figure illustrates a clear segregation in 
the mean emission-weighted metallicities for CF and NCF clusters. }
\end{figure}

The metallicity measurements for the clusters are summarized in the Table.
For the CF clusters  we list the results obtained both with spectral
models A and C ({\it i.e.} with and without the emission from the 
cooling flows accounted for in the spectral analysis). 
For the NCFs only the results for model A are listed.

The results listed in the Table demonstrate a clear segregation between
the metallicities of the CF and NCF clusters. For the CF systems, 
the mean metallicity
determined with spectral model A is 0.34. For the NCFs, the mean value
determined with the same spectral model is 0.21. The application of a
Students t-test (accounting for the possibility of unequal variances 
in the two distributions; Press \etal 1992) indicates that the mean metallicities
for the CF and NCF clusters (determined using spectral model A) differ at 
$\gg 99.9$ per cent confidence. (The probability that the means of the two distributions are equal is $1.1
\times 10^{-5}$.) Note, however, that an F-test (Press \etal 1992) 
indicates the variances of the two distributions to be consistent at 
the 91 per cent confidence level.

For the CF clusters, the metallicities determined with spectral
model C, which incorporates the cooling-flow component, are 
slightly higher (${\bar{Z}_{\rm CF}}=0.38$) than those inferred with spectral model A. 
The use of the cooling-flow model, where appropriate, therefore only
enhances the discrepancy between the metallicities of the CF and
NCF clusters. (The probability that the mean metallicities for
the CF clusters, determined with spectral model C, and for the
NCFs, determined with spectral model A, are equal is only $4.3 \times
10^{-7}$). We note that the use of spectral model B in place of model A 
({\it i.e.} allowing the absorbing column density acting on the clusters 
to be a free parameter in the fits) leads to very similar 
results. 

The mean metallicity for the whole sample of 
30 clusters, determined with spectral model A, is $\bar{Z}=0.303\pm0.085$. 
If spectral model C is used for the CF clusters, the over all sample mean rises 
to $\bar{Z}=0.329\pm0.102$. The mean metallicity measured with model A is 
in good agreement the value of $\bar{Z}=0.27\pm0.15$ determined by Edge \& Stewart (1991), 
from EXOSAT observations of low-redshift clusters 
(using a similar spectral model. This value has been
adjusted to account for the different Fe/H ratio assumed in that study.) 
For individual clusters, our results also show good agreement 
with those of Yamashita (1992), from GINGA observations, after correction for 
the different definition of solar metallicity used in that work.
The standard deviations in the metallicity measurements for the subsamples 
of CF and NCF clusters ($5.7 \times 10^{-2}$ and $6.5 \times 10^{-2}$
for the CFs with models A and C, respectively, and $5.4 \times 10^{-2}$ for the
NCFs with model A) are smaller, by a factor of $\sim 2$, than that
for the sample as a whole, and are more in line with expectations from 
theoretical models (\eg Kauffmann \& Charlot 1997).

We note that Abell 2319 is the only NCF-classified cluster with a 
metallicity comparable to that of a CF system. This is probably 
due to our conservative approach in classifying systems as NCFs. 
Abell 2319 is the only NCF in the Table with a mean cooling time within 
100~kpc of the cluster centre of less than $10^{10}\yr$. 
Further observations will clarify the identification of Abell 2319 
as a CF or NCF cluster.

\begin{figure}
\centerline{\hspace{3cm}\psfig{figure=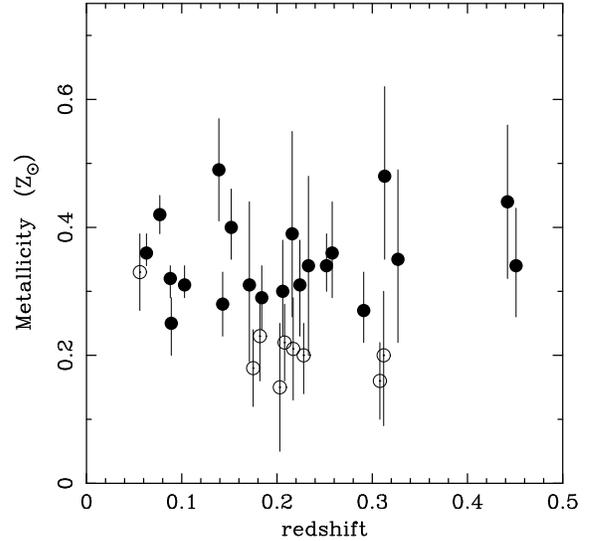,width=0.65\textwidth
,angle=270}}
\caption{The metallicity measurements for the clusters (determined using spectral
model A) as a function of redshift. The figure illustrates a lack of 
evolution in the emission-weighted metallicities of clusters within $z
\sim 0.3$.}
\end{figure}

Sixteen of the clusters included in our study (11 CFs and 5 NCFs) are
also included in the study of Mushotzky \& Lowenstein (1997). 
However, these authors only included a cooling-flow component in the
spectral analysis for two of the eleven CF clusters in their
sample (Zwicky 3146 and Abell 2390). In general, our results 
show good agreement with those of Mushotzky \& Lowenstein (1997), although our values 
for a few of the CF clusters are slightly higher. This is presumably due 
to recent improvements in the calibration of the the ASCA instruments.

We have searched for evidence of evolution in the cluster metallicities by
splitting the clusters into two subsamples, with redshifts less 
and greater than the mean value of $z=0.210$. This gives 16 clusters with
$z<0.210$ (mean redshift 0.14) and 14 with $z>0.210$ (mean redshift 0.29), 
with an approximately even mix of CF and
NCF systems in the two samples (5 and 4 NCFs 
in the low and high redshift samples, respectively). The mean
metallicities for the low$-z$ and high$-z$ samples (using spectral model A
for all clusters) are $0.30\pm0.08$ and $0.31\pm0.09$, respectively.
The application of the Students t-test 
shows the difference between the mean metallicities for the high$-z$ and
low$-z$ samples to be significant at only the 18 per cent confidence 
level. In agreement with Mushotzky and Lowenstein (1997) we therefore 
find no evidence for evolution in the emission-weighted metallicities over
the range of redshifts covered by our study. We also note that we find no
evidence for significant evolution in the X-ray gas mass fraction, with
a mean value for both the low and high redshift subsamples of 0.16. 

\section{Discussion} 

\subsection{Metallicity gradients, mergers and mixing} 

Fabian \etal (1994a) suggested that the higher emission-weighted 
metallicities for CF clusters could be due to inhomogeneities in the ICM, 
with small blobs of cooler, denser and more 
metal-rich gas being immersed in hotter gas at the centres of the CF
systems. The distributed mass deposition profiles within cooling
flows requires that the ICM there is inhomogeneous (Fabian 1994 and
references therein). Fabian \etal (1994a) also noted the potential importance of
abundance gradients in clusters. Abundance gradients appear to be a common 
feature of the cores of CF clusters. First noticed in ASCA (Fukazawa \etal 1994) 
and ROSAT PSPC (Allen \& Fabian 1994) spectra of the Centaurus cluster, 
an increase in the iron abundance towards the centres of CF clusters has
since been found in ASCA spectra for Abell 496 (Hatsukade 1997), the Virgo
cluster (Matsumoto \etal 1996) and AWM7 (Ezawa \etal
1997). The abundance of iron across the Perseus cluster, which also hosts a large 
cooling flow, appears patchy with the highest value probably occurring 
in the core (Arnaud \etal 1994; see also Molendi \etal 1998). In contrast, ASCA studies of Abell 
1060 (Tamura \etal 1996), the Ophiuchus cluster (Matsuzawa \etal 1996) 
and the Coma Cluster (Watanabe \etal 1997), which have little or no cooling flows, 
show no abundance gradients. 

An abundance gradient will strongly influence the emission-weighted 
metallicity determined from the integrated X-ray spectrum of a cluster. 
The X-ray emission depends on the square of the ICM density and so is 
dominated by the innermost, densest regions of a cluster. This is
particularly the case for CF clusters, where a significant fraction
(up to $\sim 70$ per cent) of the total X-ray luminosity may arise from within the 
central $2-3$ hundred kpc (Peres \etal 1997; Allen \etal 1998). 
We have simulated the effects of metallicity gradients on the mean
emission-weighted metallicities for CF and NCF clusters using the
observed X-ray surface brightness profiles for Abell 478 and 2218 as 
representative examples of CF and NCF systems. (These clusters 
have the highest-quality imaging data.) We find that a linear gradient in metallicity dropping
from $\sim 0.8$ solar at the centre to $\sim 0.1$ solar at 500 kpc (and remaining 
roughly constant outside this radius) leads to a mean emission-weighted 
metallicity, for a CF cluster like Abell 478, of about 0.4. 
For a NCF system, for which the X-ray surface brightness profile
will be much less sharply-peaked, the mean emission-weighted metallicity 
will be only  $\sim 0.2$. Such gradients should be easily detectable with 
AXAF.

The abundance gradient model provides the most natural explanation for
the segregation in the metallicity results for the CF and NCF clusters.
The metal-rich core may be the oldest part of a cluster, 
forming earliest in the deep potential wells. This could enhance both 
the formation of massive stars and the retention of gas in these regions.
(We note that many of the most-massive CF clusters 
also show evidence for ongoing star formation in their cores; \eg Allen 1995).  
After their initial collapse, clusters continue to evolve by the accretion of 
material, often via subcluster merger events. NCF clusters, such
as those included in this study, are thought to have recently 
experienced a major merger event wherein a large mass component 
has strongly interacted with the cluster core (Allen 1998). 
Such events will significantly disrupt the X-ray gas in the core regions 
of the clusters and will mix and 
spread the central high-metallicity gas 
with the outer less metal-rich material. The abundance gradients in 
NCF clusters are thereby reduced (or even destroyed), although the {\it total} 
mass of metals in the ICM is unchanged. 
(An implication of this 
is that the global metallicity of the ICM is 
more accurately estimated for the NCF 
systems, where the metals are more evenly mixed with the cluster gas.) 
The dependence of cluster metallicities
on the presence or absence of a cooling flows then reflects 
whether these systems have had their core regions left
undisturbed, or recently mixed, rather than on any internal
property of the cooling flows such as their mass deposition rates.

Reisenegger \etal (1996) have suggested that 
the abundance gradient in the Centaurus cluster could result from the 
cooling flow in that cluster concentrating the 
metals ejected by type-Ia supernovae in the outer regions
of the cD galaxy. Although this mechanism is plausible 
for low-luminosity systems like the Centaurus cluster, with
relatively high ratios of the stellar mass in the central galaxy 
to the X-ray gas mass, it has more difficulty in accounting for the 
gradients in more X-ray luminous clusters, 
with lower stellar/X-ray gas mass ratios. 
This implies that the metallicity gradients in luminous CF
clusters may have been present since some early epoch in their 
formation history.

A further possibility to explain the metallicity gradients in 
cooling-flow clusters is that most of the metals in cluster cores 
may reside in large grains. The lifetime of grains of radius $a\mu$m
to sputtering in hot gas of density $n$ is $\sim
2\times 10^6 a/n\yr$ (Draine \& Salpeter 1979). Provided that individual grains 
exceed 10$\mu$m in radius, they should survive for a Hubble time or longer throughout
NCF objects and beyond the cooling radius in clusters with cooling flows. 
Within cooling flows, the density rises inward so the grains are increasingly 
sputtered, releasing the metals into the gas phase which thus 
becomes increasingly metal rich toward the cluster centre. Such a
model requires that about half the metals are originally injected into the
ICM as large grains. We note that large grains are inferred
to have formed and to carry most of the iron in the expanding remnant of
SN1987A (Colgan \etal 1994). Within this model the typical mean
metallicity in the {\it core} of a cluster would be 0.5--1.0 solar. 
The grains would also provide a possible source
for the dust inferred in the central optical nebulosities in cooling flows
(Fabian, Johnstone \& Daines 1994b; Voit \& Donahue 1995; Allen \etal
1995) and may be related to the excess soft X-ray absorption observed in
cooling flows (\eg Allen \& Fabian 1997).

\subsection{The effects of cluster evolution} 

Within standard formation scenarios, clusters that form earlier are
expected to have higher luminosities for a given temperature (since the
gas density is higher at earlier formation epochs). Scharf \& Mushotzky
(1997) presented results from a single-temperature 
analysis of $\sim 30$ clusters observed with ASCA and demonstrated 
a positive correlation between the amplitude,
$A_{\rm LT}$, of the $L_{\rm X}-T_{\rm X}$ relation 
(where they define $L_{\rm X} = A_{\rm LT}T_{\rm X}^3$) and the 
mean (emission-weighted) metallicity. They suggest that the origin of this 
correlation is that clusters with larger values of $A_{\rm LT}$ formed 
earlier and were better able 
to hold on to the metal-enriched gas expelled from their  galaxies. 
Assuming that 
metals cannot be lost from the cluster potentials without a corresponding 
decrease in the X-ray gas mass, clusters with lower $A_{\rm LT}$ should 
have lower baryon fractions by about a factor of two, given the observed 
range in metallicities. 

The Table lists the baryon fractions at a radius of 500 kpc 
in the clusters, determined with the appropriate spectral models 
(spectral model C for the CF systems and model A for the NCFs). We see that 
the mean baryon fractions for the subsamples of CF and NCF clusters 
differ by only $\sim 30$ per cent, and thus that relatively little gas 
has escaped the potentials of the NCF clusters relative to the CF systems. 
(We note, however, that the distributions of baryon fraction values for the 
CF and NCF clusters are different. The application of a Student's t-test 
shows the mean values for the two subsamples to differ at 
$>99$ per cent confidence. The NCF clusters also 
have a significantly smaller 
dispersion in baryon fractions than the CF systems, and a Kolmogorov-Smirnov test 
shows the two subsamples to be drawn from different 
populations at $>99$ per cent significance.) If spectral model B 
rather than model A is used for the NCF systems ({\it i.e.} if the absorbing 
column density is included as a free parameter in the spectral analysis of 
the NCF systems) the mean baryon fraction for these clusters rises to 0.16,
improving the agreement with the CF clusters. 

It is important to note that the clusters studied in this Letter are
amongst the most X-ray luminous and by implication  most-massive clusters known. 
Material expelled from their member galaxies, 
even when part of an early low-mass subclump,
is unlikely to escape from the total cluster potential.
The accompanying Letter (Allen \& Fabian 1998) discusses 
the impact of cooling flows on the $L_{\rm Bol}-T_{\rm X}$ 
relation for clusters. Allen (1998) discusses the effects of cooling
flows on X-ray mass measurements. The new data presented 
in this Letter reveal a link between cooling flows and metallicity 
measurements and suggest the presence of metallicity gradients 
in clusters with cooling flows. 
Such effects must be accounted for before attempting to determine 
cosmological parameters from X-ray 
observations of clusters.

\section*{Acknowledgments}

We thank the Royal Society for support.

\end{document}